# Magnetic Oscillation of Optical Phonon in ABA- and ABC-Stacked Trilayer Graphene

Chunxiao Cong,[1] Jeil Jung,[2,3,4] Bingchen Cao,[1] Caiyu Qiu,[1] Xiaonan Shen,[1] Aires Ferreira,[3,4,5] Shaffique Adam,[3,4] and Ting Yu[1,3,4,†]

[1]Division of Physics and Applied Physics, School of Physical and Mathematical Sciences, Nanyang Technological University, 637371, Singapore; [2]Department of Physics, University of Seoul, Seoul, 130-743, Korea; [3]Department of Physics, Faculty of Science, National University of Singapore, 117542, Singapore; [4]Graphene Research Centre, National University of Singapore, 117546, Singapore; [5]Department of Physics, University of York, YO10 5DD, United Kingdom

[†]Address correspondence to yuting@ntu.edu.sg

## ABSTRACT

We present a comparative measurement of the G-peak oscillations of phonon frequency, Raman intensity and linewidth in the Magneto-Raman scattering of optical $E_{2g}$ phonons in mechanically exfoliated ABA- and ABC-stacked trilayer graphene (TLG). Whereas in ABA-stacked TLG, we observe magnetophonon oscillations consistent with single-bilayer chiral band doublets, the features are flat for ABC-stacked TLG up to magnetic fields of 9 T. This suppression can be attributed to the enhancement of band chirality that compactifies the spectrum of Landau levels and modifies the magnetophonon resonance properties. The drastically different coupling behaviour between the electronic excitations and the $E_{2g}$ phonons in ABA- and ABC-stacked TLG reflects their different electronic band structures and the electronic Landau level transitions and thus can be another way to determine the stacking orders and to probe the stacking-order-dependent electronic structures. In addition, the sensitivity of the magneto-Raman scattering to the particular stacking order in few layers graphene highlights the important role of interlayer coupling in modifying the optical response properties in van der Waals layered materials.

## I. INTRODUCTION

The physics of single and few layers graphene has emerged as a prolific research area during the last decade.[1-3] Trilayer graphene (TLG) bridges the single atomic layer and the bulk graphite limits and its crystalline form can display Bernal (ABA) and rhombohedral (ABC) stacking orders.[4-10] The TLG's stacking-dependent band structures near the Dirac point has been assessed by several methods including integer quantum Hall effect (IQHE) measurements,[8] infrared (IR) absorption spectroscopy,[10] scanning tunneling microscopy (STM),[11] angle-resolved photoemission spectroscopy (ARPES),[12] and Raman spectroscopy.[13-15] Among these, Raman spectroscopy is a commonly used method owing to its unique advantages of processing simplicity, high efficiency, non-destructiveness and reliability, and has been widely employed to probe properties of graphene layers, such as the electronic[16-19] and crystal[14, 20-22] structures, and optical,[23, 24] mechanical,[25-27] electrical doping,[28, 29] and chemical functionalization properties.[30-32] Moreover, Raman spectroscopy is also widely used to investigated the magneto-phonon resonance (MPR) effect which induced by the strong coupling of the inter-Landau level (LL) electronic transitions with the doubly degenerate $E_{2g}$ optical phonons in graphene layers and bulk graphite under magnetic fields, which gives important information on electron-phonon interactions and the transitions between quantized Landau levels.[33-40] The MPR in monolayer graphene and bilayer graphene has been theoretically predicted.[41-44] Experimentally, the MPR effect was observed in different types of graphene samples such as decoupled monolayer graphene on graphite, monolayer graphene and Bernal stacked few-layer graphene on $SiO_2$/Si and non-Bernal stacked multilayer graphene on SiC.[33, 34, 36, 38, 39, 45-47] However, a comparative study of stacking-order-dependent MPR effect in ABA- and ABC-stacked TLG has not been reported in previous works.

In this article, we investigate the phonon-electron coupling in TLG under the influence of a magnetic field, and the role played by the layer stacking, using magneto-micro-Raman spectroscopy. It is found that the $E_{2g}$ phonon shows drastically different behaviors for ABA- and ABC-stacked TLG under magnetic fields. In addition to the changes of $E_{2g}$ phonon energies with varying magnetic fields, we observe well-defined oscillations in the Raman intensity and linewidth of the $E_{2g}$ phonons in both supported and suspended ABA-stacked TLG. This *genuine* MPR effect presented in both supported and suspended samples indicates a strong coupling of the electronic Landau level transitions and optical phonons and its weak dependence on the substrate. In contrast, the phonon energy, the Raman intensity, and the linewidth of the $E_{2g}$ phonons of ABC-stacked TLG have no features up to magnetic fields of 9 T. The quenching of the MPR features in ABC-stacked TLG below 9 T can be related with the tighter bunching of the Landau levels resulting in higher chirality bands. The remarkably different coupling behaviours between electronic excitations and $E_{2g}$ phonons in ABA- and ABC-stacked TLG reflect the magnetophonon response sensitivity to the interlayer coupling details.

## II. EXPERIMENT

Suspended and supported graphene layers were prepared by micromechanical cleavage[1] of natural graphite crystals and deposited on 300 nm $SiO_2$/Si substrates with and without pre-patterned holes. The detailed fabrication process of the suspended samples is described in our previous work.[48] An optical microscope was used to locate the thin layers, and the number of layers was further identified by white light contrast spectra and the absolute Raman intensity of the G mode at room temperature.[49] The stacking sequence was then determined by the linewidth of the Raman G'(2D) mode at room temperature.[13, 14] The white light contrast spectra were acquired using a WITec CRM200 Raman system with a 150 lines/mm grating. The room temperature Raman spectra\images were obtained using a WITec CRM200 Raman system with a 600 lines/mm grating and using 532 nm laser excitation with laser power below 0.1 mW on the sample surface to avoid laser-induced heating. The low-temperature magneto-Raman spectra of the TLG samples were measured with a customer designed confocal micro-Raman spectroscopy/image system at $T$ = 4.2 K. The sample was mounted on a non-magnetic piezocrystal controlled stage consisting of *xyz* positioners with movement range of 5 mm × 5 mm × 2.5 mm under objective lens and *xy* scanners with scanning range up to 30 × 30 μm$^2$ inside a vacuum stick. After the vacuum stick was pumped to 10$^{-5}$ mbar and then filled with 20 mbar helium gas, the sample together with the vacuum stick were immersed into a cryostat filled with liquid helium. A superconducting magnet was installed inside the cryostat, delivering magnetic fields up to 9 T perpendicular to the plane of graphene layers. Laser excitation at wavelength of 532 nm with a ~ 5 mW laser power was introduced by a 5 μm core mono-mode optical fiber onto the sample. The laser spot on the sample was ~ 1 μm in diameter. The scattered light was collected by a multimode optical fiber with a 100 μm core in the backscattering configuration and then was dispersed by a single grating spectrometer (1800 lines/mm) equipped with a thermoelectrically cooled CCD. The magneto-Raman measurements were performed at a step of 0.1 T with the magnetic field.

## III. EXPERIMENTAL RESULTS AND DISCUSSIONS

Typical room temperature Raman spectra/images of mechanically exfoliated graphene sheets with different number of layers together with an optical image and a white light reflectance image, are presented in Fig. 1. The homogeneous contrast of the optical image (Fig. 1(a)), the white light reflectance image (Fig. 1(c)), and the Raman image of the G band intensity (Fig. 1(d)) for the area highlighted by light blue dashed line clearly shows that the sample thickness or the number of layers of the highlighted part is identical. It can be determined by the extracted contrast value and Raman spectra (Fig. 1(b)) that the highlighted part is TLG. However, there are two parts with strikingly different contrasts in the Raman image extracted from the G'(2D) band linewidth of the TLG region as shown in Fig. 1(e). This indicates the existence of two domains with different stacking orders. Using the double resonance Raman theory and considering the unique electronic band structures of ABA- and ABC-stacked TLG especially at the low energy level, we can assign the brighter (broader G' mode) part as ABC-stacked TLG, while the darker (narrower G' mode) part is assigned as ABA-stacked TLG as discussed previously.[13, 14]

To investigate the inter Landau level excitations (or magnetoexcitons between LLs) and their interactions with phonons of ABA- and ABC-stacked TLG under magnetic field, which can further probe the energy dispersion spectra of TLG with different stacking order, low-temperature magneto-Raman scattering measurements in the spectral range of the vicinity of the $E_{2g}$ optical phonon energy were performed on our ABA- and ABC-stacked TLG at $T$ = 4.2 K under $\lambda$ = 532 nm laser excitation. The representative magneto-Raman spectra of the ABA- and ABC-stacked TLG at certain values of magnetic field are displayed in Fig. 4 (see Appendix A). No direct observation of electronic excitations in the Raman spectra of neither ABA- nor ABC-stacked TLG. All the magneto-Raman spectra under magnetic fields from 0 T to 9 T of both ABA- and ABC-stacked TLG keep single Lorentzian line shapes. No distinct split peaks of the G mode are observed. We show the magneto-Raman spectra of ABA- and ABC-stacked TLG in the form of a false-colour map of the Raman intensity as a function of the magnetic field in Figs. 2(a) and 2(b). The Raman intensities are shown by colour scales. Regarding ABA-stacked TLG (see Fig. 2(a)), in addition to the change of the frequency of the $E_{2g}$ phonon with varying magnetic field, we observe a clear variation in the Raman intensity of the $E_{2g}$ phonon. However, both the frequency and the Raman intensity of the $E_{2g}$ phonon of ABC-stacked TLG are inert to the magnetic field as shown in Fig. 2(b). To clearly show the detailed magnetic-field-dependent evolutions of the $E_{2g}$ phonon of ABA- and ABC-stacked TLG, the peak position, full width at half maximum intensity (FWHM), and peak height of the $E_{2g}$ phonon of ABA- and ABC-stacked TLG, which are obtained by fitting the Raman G band at different values of magnetic field as a single Lorentzian peak, are shown in Figs. 2(c)-2(e). It can be seen from Figs. 2(c)-2(e) that distinctive stacking-dependent features are observed. The G peak position, FWHM, and peak height of ABA-stacked TLG oscillate with magnetic field, while they are inert in ABC-stacked TLG up to 9 T. The two most prominent resonant magnetic fields for ABA-stacked TLG are at around 2.5 T and 4.5 T, which can also be seen clearly from Figs. 2(c), 2(b) and 2(e). A comparative study of MPR effects in both supported and suspended ABA-stacked TLG reveals the substrate has no significant effects on the magneto-phonon resonances of ABA-stacked TLG (see more details in Appendix B and Fig. 5). Furthermore, the fast Fourier transforms (FFTs) of the G peak's position, FWHM, and height in ABA-stacked TLG signal out a unique frequency at around 6.0 T [see Figs. 2(f)–2(h)], which represents an inverse period of ABA-stacked TLG. By considering its electronic spectrum, consisting of monolayer-type bands and bilayer-type bands, and the condition for magneto-phonon resonance, the calculated values of inverse periods of ABA-stacked TLG are 7.0 T and 36.5 T, respectively (see more details in Appendix D). The characteristic frequency in the FFT signals at 6.0 T is close to the calculated value of 7.0 T associated with the monolayer-type band.

Now we elucidate the experimental observations of the distinctive stacking-dependent magnetophonon oscillation features in ABA- and ABC-stacked TLG by numerical calculations. We note that the observed oscillatory features would manifest as anticrossing branches in clean enough devices.[34] The magneto-Raman $E_{2g}$ phonon frequency oscillations are related with the inter-Landau-level resonant transitions $T_n$ between states with Landau level index differences $|n| - |m| = \pm 1$ from the initial filled $m$ to final $n$ empty levels. The filling-factor dependent magnetophonon frequency renormalization given in Ref. [43, 50] reduces to a simpler form in a charge neutral system

$$\tilde{\varepsilon}^2 - \varepsilon_0^2 = 2\lambda_{ep}\varepsilon_0 \left( t + E_1^2 \sum_{n=0}^{\infty} \frac{T_n}{\tilde{\varepsilon}^2 - T_n^2} \right), \tag{1}$$

where the phonon self-energy uses the bare propagator[43, 50] and assumes the same form of magnetic field dependence for the electron-phonon coupling $\propto B^{1/2}$. We defined $\tilde{\varepsilon} = \epsilon + i\delta$. Here, ε is the renormalized phonon frequency, $\delta$ is the broadening, $\varepsilon_0$ is the dressed phonon frequency measured at zero field, $\lambda_{ep}$ is the dimensionless electron-phonon coupling constant, $E_1 = v_F(2\hbar eB)^{1/2}$ is used to capture the $B$-field evolution of the electron-phonon coupling, and the parameter $t$ that physically is related with the nearest neighbour hopping term accounts for the difference between the bare phonon frequency $\varepsilon_{ph}$ and $\varepsilon_0$ through the relation $\varepsilon_{ph}^2 = \varepsilon_0^2 + 2\varepsilon_0\lambda_{ep}t$. Considering the uncertainty in the microscopic models for electron-phonon coupling, in the magnetophonon resonance calculations this term can also be interpreted as a phenomenological parameter. In the study of neutral Dirac-like system of multilayer epitaxial graphene grown on SiC by Faugeras *et al.*,[47] the approximation $t \simeq \sum_{n=1}^{N_c} E_1^2 / T_n$ was used with the sum truncated at a given LL cutoff $N_c$. In our calculations, $t$ is an adjustable parameter to shift the origin of the renormalized dressed frequencies to partly account for the cutoff effects in the fixed number of one thousand LLs in the sum. The inter Landau level transition energies $T_n = |E_n - E_m|$ between the $n^{th}$ level and the respective counterpart allowed by the selection rule $|n| - |m| = \pm 1$ apply for $m \to n$ inter LL optical transitions.

We use Eq. (1) to interpret the resonant phonon frequency renormalization using values of $T_n$ corresponding to ABA- or ABC-stacked TLG. In the case of ABA-stacked TLG whose bands consist of a sum of a monolayer-like and a bilayer-like dispersion, we notice that the magnetophonon oscillation is almost entirely captured by the monolayer-like oscillation whereas the bilayer part has a marginally small contribution to the magnetophonon oscillation. We used the same parameters as in Ref. [47] where $v_F = 1.02 \times 10^6$ m/s, $\lambda_{ep} \sim 4.5 \times 10^{-3}$, $\varepsilon_0 = 1586.5$ cm$^{-1}$ but two and a half times stronger disorder broadening of $\delta = 2.5 \times 90$ cm$^{-1}$ to obtain magnetophonon oscillation amplitudes in ABA-stacked TLG close to those observed in experiments, see Fig. 3. The experimental magnetophonon features of ABC-stacked TLG are completely flat suggesting that the suppression of the oscillations can be related with the closer spacing of the Landau level spectra in the bands with high chirality. This behavior is apparent from the simplified low energy expression of the Landau level energies in ABC-stacked TLG $E_n = \xi\left((2v_F^2/l_B^2)^{3/2}/t_\perp^2\right)(n(n+1)(n+2))^{1/2}$ [51] where we use $t_\perp = 0.36$ eV, the effective interlayer coupling strength obtained for bilayer in the Local Density Approximation.[52] In the low field limit, we can see that the higher power law scaling of the *B*-field results in a smaller leading coefficient and therefore more closely spaced energy levels which makes the resonant transitions more susceptible to the effects of broadening. We introduce an effective parameter $\xi$ to control the separation of the Landau levels. The value $\xi = 1$ in the low energy form is valid to approximate the minimal model full bands spectrum for small fields but the errors in the dispersion for moderately large *B*-fields of a few Teslas become substantial due to the inaccurate $B^{3/2}$ scaling.[53] Thus, this parameter is meant to reduce the overall width of the Landau level spectrum spacing to correct for this spurious separation of the simplified Landau levels at moderately large fields. By using a smaller value $\xi = 1/4$, we reduce the spacing between the Landau levels that leads to a separations similar to the full bands calculation when B~9 T for the first twenty Landau levels, see Fig. 3. Since both the ABA- and ABC-stacked TLG are simultaneously present in the same sample experimentally, we use the same value of the disorder broadening δ in our calculations. The resulting magnetophonon oscillations for ABC-stacked TLG in Fig. 3 clearly show that the closer spacings between the Landau levels lead to suppressed magnetophonon oscillations due to the smearing of the features by the disorder broadening.

## IV. CONCLUSIONS

In summary, we have investigated the low temperature (~ 4.2 K) magneto-Raman scattering response of $E_{2g}$ phonons in ABA- and ABC-stacked TLG under magnetic fields up to 9 T. Noticeable MPR effects are observed in ABA-stacked TLG but not in ABC-stacked TLG under magnetic fields up to 9 T, signalling a clear stacking order dependence of MPR effects in TLG. The shapes of the MPR in ABA-stacked TLG, whose band structure can be decomposed into that of a monolayer and bilayer graphene, showed mainly features of a monolayer graphene for the magnetic field ranges explored indicating that the oscillations due to higher chirality bilayer bands appear only as small perturbations to the monolayer features. A comparative study of MPR effects in both supported and suspended ABA-stacked TLG reveals a weak substrate-dependence. The absence of the MPR effects in ABC-stacked TLG below 9 T was attributed to the closer energy spacing between the Landau levels that makes the magnetophonon oscillations more susceptible to disorder broadening. The experimental observations were supported by numerical calculations of the magnetophonon oscillations that assume the same value of disorder broadening but uses as input different Landau level spectra that correspond to the ABA and ABC layers. Our results strongly suggest that magnetophonon oscillation probed in multilayer graphene or graphite will predominantly show features of monolayer graphene over bilayer graphene or higher band chirality rhombohedral multilayer. Thus it is expected that probing magneto-Raman oscillations associated to multilayer graphene with rhombohedral stacking will normally require larger magnetic fields and samples with low disorder.

*Acknowledgements*

The work described here was supported by National Research Foundation-Prime Minister's office, Republic of Singapore (NRF-RF2010-07, NRF-NRFF2012-01, R-144-000-302-281, NRF-CRP R-144-000-295-281), and by Ministry of Education - Singapore (MOE2013–T1–2–235, MOE2012-T2-2-049). We thank A. H. MacDonald for providing insights and pointing to the key references and V. Pereira and J. Viana-Gomes for helpful discussions.

**APPENDIX A: REPRESENTATIVE MAGNETO-RAMAN SPECTRA OF ABA- AND ABC-STACKED TLG**

Figure 4 displays the representative Raman spectra of ABA- and ABC-stacked TLG at different values of magnetic field measured at $T = 4.2$ K under $\lambda = 532$ nm laser excitation. It can be seen that the $E_{2g}$ phonon shows drastically different behaviour for ABA- and ABC-stacked TLG though there is no direct observation of electronic excitations in the Raman spectra of neither ABA- nor ABC-stacked TLG.

In Fig. 4(a) of ABA-stacked TLG, the most prominent resonant magnetic fields for ABA-stacked TLG are observed between 4 T and 5 T, as indicated by the two green dashed curves, while in Fig. 4(b) of ABC-stacked TLG, no changes are observed in the magnetic field range of measurement.

**APPENDIX B: COMPARISON BETWEEN THE TRSULTS OF MAGNETO-RAMAN MEASUREMENTS ON SUSPENDED AND SUPPORTED ABA-STACKED TLG**

To investigate the substrate effects on the observed magnetophonon resonances in ABA-stacked TLG, suspended and $SiO_2$/Si-substrate supported ABA-stacked TLG samples were prepared and located. The corresponding magnetic field evolutions of the G peak position and the G peak FWHM of both suspended and supported ABA-stacked TLG are shown in Fig. 5. The phenomena of magnetophonon resonances in both suspended and supported ABA-stacked TLG observed are almost the same as the results of ABA-stacked TLG displayed in Fig. 2. Therefore, it can be concluded that the substrate has no significant effects on the magnetophonon resonances of ABA-stacked TLG.

**APPENDIX C: MAGNETO-RAMAN RESULTS OF A SECOND ABC-STACKED TLG SAMPLE**

The magneto-Raman measurements on a second ABC-stacked TLG sample is shown in Fig. 6. The number of layers (TLG) and the stacking orders (ABA and ABC) of the graphene flakes marked by light blue dashed line in the optical image of Fig. 6(a), are differentiated by white light reflectance image and Raman images (see Fig. 6(b-d)). The magneto-Raman results of the ABA-stacked region (not shown here) are almost the same as the phenomena we observed in Fig. 2 of ABA-stacked TLG.

However, the tiny fluctuations in the magnetic field range of 1-3 T appeared in Fig. 2(c-e) of ABC-stacked TLG did not reproduce in this ABC-stacked region, as shown in Fig. 6(e-g). Therefore, the fluctuation should not relate to MPR effect.

**APPENDIX D: CALCULATION OF THE INVERSE PERIODS OF ABA-STACKED TLG**

The inverse periods of ABA-stacked TLG can be estimated by considering its electronic spectrum under the magnetic field consisting of a series of discrete inter-Landau levels and the condition for magneto-phonon resonance. The low-energy Landaul level spectrum of ABA-TLG consists of monolayer-type bands ($k_z = \pi/2$) and bilayer-type bands ($k_z = \pi/4$), which are given by Eq. (A1) and (A2) below, respectively.[54]

$$E_n = \pm\sqrt{2\hbar v_F^2 eB}\sqrt{n}, \tag{A1}$$

$$E_n = \pm\frac{\hbar v_F^2 eB}{t_\perp \cos k_z}\sqrt{n(n+1)}, \tag{A2}$$

where $v_F$ is the Fermi velocity in the monolayer and $t_\perp$ is the nearest-neighbor inter-layer hopping amplitude. From now on, we use units such that $\hbar \equiv 1 \equiv c$. The inter Landau level excitation energies for inter-band transitions of $-(n-1) \rightarrow n$ for monolayer-type bands and bilayer-type bands are then given by

$$\Delta E = \sqrt{2\hbar v_F^2 eB}(\sqrt{n} + \sqrt{n-1}), \tag{A3}$$

$$\Delta E = \frac{\hbar v_F^2 eB}{t_\perp \cos k_z}(\sqrt{n(n+1)} + \sqrt{n(n-1)}), \tag{A4}$$

In equations of (A3) and (A4), we assume $n \gg 1$ (approximating $\sqrt{n(n\pm 1)} \approx n \pm 1/2$), and consider the magneto-phonon resonance condition of $\Delta E = \hbar\omega_{\text{ph}}$, then we obtain

$$\frac{1}{B_n} = \frac{8v_F^2 e}{\omega_{ph}^2}(n - 1/2), \tag{A5}$$

$$\frac{1}{B_n} = \frac{2v_F^2 e}{\omega_{ph} t_\perp \cos(\frac{\pi}{4})} n. \tag{A6}$$

The coefficients in front of *n* of Eq. (A5) and (A6) give the inverse periods for monolayer-type bands and bilayer-type bands of ABA-stacked TLG, respectively.[45] For $\omega_{ph} = 196$ meV, $v_F = 1.02 \times 10^6$ ms$^{-1}$, and $t_\perp = 360$ meV, solving Eq. (A5) and (A6) brings inverse periods of 7.0 T and 36.5 T, respectively.

**References**

[1] K. S. Novoselov, A. K. Geim, S. V. Morozov, D. Jiang, Y. Zhang, S. V. Dubonos, I. V. Grigorieva, and A. A. Firsov, Electric field effect in atomically thin carbon films, Science **306**, 666 (2004).

[2] K. S. Novoselov, A. K. Geim, S. V. Morozov, D. Jiang, M. I. Katsnelson, I. V. Grigorieva, S. V. Dubonos, and A. A. Firsov, Two-dimensional gas of massless Dirac fermions in graphene, Nature (London) **438**, 197 (2005).

[3] Y. B. Zhang, Y. W. Tan, H. L. Stormer, and P. Kim, Experimental observation of the quantum Hall effect and Berry's phase in graphene, Nature (London) **438**, 201 (2005).

[4] M. F. Craciun, S. Russo, M. Yamamoto, J. B. Oostinga, A. F. Morpurgo, and S. Thrucha, Trilayer graphene is a semimetal with a gate-tunable band overlap, Nat. Nanotechnol. **4**, 383 (2009).

[5] W. Bao, L. Jing, J. Velasco, Y. Lee, G. Liu, D. Tran, B. Standley, M. Aykol, S. B. Cronin, D. Smirnov, M. Koshino, E. McCann, M. Bockrath, and C. N. Lau, Stacking-dependent band gap and quantum transport in trilayer graphene, Nat. Phys. **7**, 948 (2011).

[6] F. Zhang, B. Sahu, H. K. Min, and A. H. MacDonald, Band structure of ABC-stacked graphene trilayers, Phys. Rev. B **82**, 035409 (2010).

[7] L. Y. Zhang, Y. Zhang, J. Camacho, M. Khodas, and I. Zaliznyak, The experimental observation of quantum Hall effect of l=3 chiral quasiparticles in trilayer graphene, Nat. Phys. **7**, 953 (2011).

[8] A. Kumar, W. Escoffier, J. M. Poumirol, C. Faugeras, D. P. Arovas, M. M. Fogler, F. Guinea, S. Roche, M. Goiran, and B. Raquet, Integer Quantum Hall effect in trilayer graphene, Phys. Rev. Lett. **107**, 126806 (2011).

[9] M. Koshino and E. McCann, Gate-induced interlayer asymmetry in ABA-stacked trilayer graphene, Phys. Rev. B **79**, 125443 (2009).

[10] A. A. Avetisyan, B. Partoens, and F. M. Peeters, Stacking order dependent electric field tuning of the band gap in graphene multilayers, Phys. Rev. B **81**, 115432 (2010).

[11] A. G. S. Hattendorf, M. Liebmann, and M. Morgenstern, Networks of ABA and ABC stacked graphene on mica observed by scanning tunneling microscopy, Surf. Sci. **610**, 53 (2013).

[12] C. Coletti, S. Forti, A. Principi, K. V. Emtsev, A. A. Zakharov, K. M. Daniels, B. K. Daas, M. V. S. Chandrashekhar, T. Ouisse, D. Chaussende, A. H. MacDonald, M. Polini, and U. Starke, Revealing the electronic band structure of trilayer graphene on SiC: An angle-resolved photoemission study, Phys. Rev. B **88**, 155439 (2013).

[13] C. H. Lui, Z. Q. Li, Z. Y. Chen, P. V. Klimov, L. E. Brus, and T. F. Heinz, Imaging Stacking Order in Few-Layer Graphene, Nano Lett. **11**, 164 (2011).

[14] C. X. Cong, T. Yu, K. Sato, J. Z. Shang, R. Saito, G. F. Dresselhaus, and M. S. Dresselhaus, Raman characterization of ABA- and ABC-stacked trilayer graphene, ACS Nano **5**, 8760 (2011).

[15] C. X. Cong, K. Li, X. X. Zhang, and T. Yu, Visualization of arrangements of carbon atoms in graphene layers by Raman mapping and atomic-resolution TEM, Sci. Rep. **3**, 1195 (2013).

[16] L. M. Malard, D. L. Mafra, S. K. Doorn, and M. A. Pimenta, Resonance Raman scattering in graphene: Probing phonons and electrons, Solid State Commun. **149**, 1136 (2009).

[17] A. C. Ferrari, J. C. Meyer, V. Scardaci, C. Casiraghi, M. Lazzeri, F. Mauri, S. Piscanec, D. Jiang, K. S. Novoselov, S. Roth, and A. K. Geim, Raman spectrum of graphene and graphene layers, Phys. Rev. Lett. **97**, 187401 (2006).

[18] L. M. Malard, J. Nilsson, D. L. Mafra, D. C. Elias, J. C. Brant, F. Plentz, E. S. Alves, A. H. C. Neto, and M. A. Pimenta, Electronic properties of bilayer graphene probed by Resonance Raman Scattering, Phys. Status Solidi B **245**, 2060 (2008).

[19] M. S. Dresselhaus, L. M. Malard, M. A. Pimenta, and G. Dresselhaus, Raman spectroscopy in graphene, Phys. Rep. **473**, 51 (2009).

[20] Y. M. You, Z. H. Ni, T. Yu, and Z. X. Shen, Edge chirality determination of graphene by Raman spectroscopy, Appl. Phys. Lett. **93**, 163112 (2008).

[21] C. X. Cong, T. Yu, and H. M. Wang, Raman study on the G mode of graphene for determination of edge orientation, ACS Nano **4**, 3175 (2010).

[22] C. X. Cong, T. Yu, R. Saito, G. F. Dresselhaus, and M. S. Dresselhaus, Second-order overtone and combination Raman modes of graphene layers in the range of 1690-2150 cm$^{-1}$, ACS Nano **5**, 1600 (2011).

[23] T. T. Tang, Y. B. Zhang, C. H. Park, B. S. Geng, C. Girit, Z. Hao, M. C. Martin, A. Zettl, M. F. Crommie, S. G. Louie, Y. R. Shen, and F. Wang, A tunable phonon-exciton Fano system in bilayer graphene, Nat. Nanotechnol. **5**, 32 (2010).

[24] K. Kang, D. Abdula, D. G. Cahill, and M. Shim, Lifetimes of optical phonons in graphene and graphite by time-resolved incoherent anti-Stokes Raman scattering, Phys. Rev. B **81**, 165405 (2010).

[25] T. Yu, Z. H. Ni, C. L. Du, Y. M. You, Y. Y. Wang, and Z. X. Shen, Raman mapping investigation of graphene on transparent flexible substrate: The strain effect, J. Phys. Chem. C **112**, 12602 (2008).

[26] Z. H. Ni, T. Yu, Y. H. Lu, Y. Y. Wang, Y. P. Feng, and Z. X. Shen, Uniaxial strain on graphene: Raman spectroscopy study and band-gap opening, ACS Nano **2**, 2301 (2008).

[27] T. M. G. Mohiuddin, A. Lombardo, R. R. Nair, A. Bonetti, G. Savini, R. Jalil, N. Bonini, D. M. Basko, C. Galiotis, N. Marzari, K. S. Novoselov, A. K. Geim, and A. C. Ferrari, Uniaxial strain in graphene by Raman spectroscopy: G peak splitting, Gruneisen parameters, and sample orientation, Phys. Rev. B **79**, 205433 (2009).

[28] J. Yan, Y. B. Zhang, P. Kim, and A. Pinczuk, Electric field effect tuning of electron-phonon coupling in graphene, Phys. Rev. Lett. **98**, 166802 (2007).

[29] J. Yan, E. A. Henriksen, P. Kim, and A. Pinczuk, Observation of anomalous phonon softening in bilayer graphene, Phys. Rev. Lett. **101**, 136804 (2008).

[30] A. Das, S. Pisana, B. Chakraborty, S. Piscanec, S. K. Saha, U. V. Waghmare, K. S. Novoselov, H. R. Krishnamurthy, A. K. Geim, A. C. Ferrari, and A. K. Sood, Monitoring dopants by Raman scattering in an electrochemically top-gated graphene transistor, Nat. Nanotechnol. **3**, 210 (2008).

[31] Z. Q. Luo, T. Yu, K. J. Kim, Z. H. Ni, Y. M. You, S. Lim, Z. X. Shen, S. Z. Wang, and J. Y. Lin, Thickness-dependent reversible hydrogenation of graphene layers, ACS Nano **3**, 1781 (2009).

[32] N. Peimyoo, T. Yu, J. Z. Shang, C. X. Cong, and H. P. Yang, Thickness-dependent azobenzene doping in mono- and few-layer graphene, Carbon **50**, 201 (2012).

[33] J. Yan, S. Goler, T. D. Rhone, M. Han, R. He, P. Kim, V. Pellegrini, and A. Pinczuk, Observation of magnetophonon resonance of Dirac Fermions in graphite, Phys. Rev. Lett. **105**, 227401 (2010).

[34] C. Faugeras, M. Amado, P. Kossacki, M. Orlita, M. Kuhne, A. A. L. Nicolet, Y. I. Latyshev, and M. Potemski, Magneto-Raman scattering of graphene on graphite: electronic and phonon excitations, Phys. Rev. Lett. **107**, 036807 (2011).

[35] P. Kossacki, C. Faugeras, M. Kuhne, M. Orlita, A. A. L. Nicolet, J. M. Schneider, D. M. Basko, Y. I. Latyshev, and M. Potemski, Electronic excitations and electron-phonon coupling in bulk graphite through Raman scattering in high magnetic fields, Phys. Rev. B **84**, 235138 (2011).

[36] Y. Kim, Y. Ma, A. Imambekov, N. G. Kalugin, A. Lombardo, A. C. Ferrari, J. Kono, and D. Smirnov, Magnetophonon resonance in graphite: High-field Raman measurements and electron-phonon coupling contributions, Phys. Rev. B **85**, 121403 (2012).

[37] M. Kuhne, C. Faugeras, P. Kossacki, A. A. L. Nicolet, M. Orlita, Y. I. Latyshev, and M. Potemski, Polarization-resolved magneto-Raman scattering of graphenelike domains on natural graphite, Phys. Rev. B **85**, 195406 (2012).

[38] C. Y. Qiu, X. N. Shen, B. C. Cao, C. X. Cong, R. Saito, J. J. Yu, M. S. Dresselhaus, and T. Yu, Strong magnetophonon resonance induced triple G-mode splitting in graphene on graphite probed by micromagneto Raman spectroscopy, Phys. Rev. B **88**, 165407 (2013).

[39] X. Shen, C. Qiu, B. Cao, C. Cong, W. Yang, H. Wang, and T. Yu, Electrical field tuning of magneto-Raman scattering in monolayer graphene, Nano Res. **8**, 1139 (2015).

[40] Y. Kim, J. M. Poumirol, A. Lombardo, N. G. Kalugin, T. Georgiou, Y. J. Kim, K. S. Novoselov, A. C. Ferrari, J. Kono, O. Kashuba, V. I. V. I. Fal'ko, and D. Smirnov, Measurement of filling-factor-dependent magnetophonon resonances in graphene using Raman spectroscopy, Phys. Rev. Lett. **110**, 227402 (2013).

[41] T. Ando, Magnetic oscillation of optical phonon in graphene, J. Phys. Soc. Jpn. **76**, 024712 (2007).

[42] T. Ando, Anomaly of optical phonons in bilayer graphene, J. Phys. Soc. Jpn. **76**, 104711 (2007).

[43] M. O. Goerbig, J. N. Fuchs, K. Kechedzhi, and V. I. Fal'ko, Filling-factor-dependent magnetophonon resonance in graphene, Phys. Rev. Lett. **99**, 087402 (2007).

[44] O. Kashuba and V. I. Fal'ko, Role of electronic excitations in magneto-Raman spectra of graphene, New J. Phys. **14**, 105016 (2012).

[45] C. Faugeras, P. Kossacki, A. A. L. Nicolet, M. Orlita, M. Potemski, A. Mahmood, and D. M. Basko, Probing the band structure of quadri-layer graphene with magneto-phonon resonance, New J. Phys. **14**, 095007 (2012).

[46] S. Berciaud, M. Potemski, and C. Faugeras, Probing electronic excitations in mono- to pentalayer graphene by micro magneto-Raman spectroscopy, Nano Lett. **14**, 4548 (2014).

[47] C. Faugeras, M. Amado, P. Kossacki, M. Orlita, M. Sprinkle, C. Berger, W. A. de Heer, and M. Potemski, Tuning the electron-phonon coupling in multilayer graphene with magnetic fields, Phys. Rev. Lett. **103**, 186803 (2009).

[48] Z. H. Ni, T. Yu, Z. Q. Luo, Y. Y. Wang, L. Liu, C. P. Wong, J. M. Miao, W. Huang, and Z. X. Shen, Probing charged impurities in suspended graphene using Raman spectroscopy, ACS Nano **3**, 569 (2009).

[49] Z. H. Ni, H. M. Wang, J. Kasim, H. M. Fan, T. Yu, Y. H. Wu, Y. P. Feng, and Z. X. Shen, Graphene thickness determination using reflection and contrast spectroscopy, Nano Lett. **7**, 2758 (2007).

[50] J. N. Fuchs, Dirac fermions in graphene and analogues: magnetic field and topological properties, arXiv:1306.0380.

[51] H. K. Min and A. H. MacDonald, Electronic structure of multilayer graphene, Prog. Theor. Phys. Supp. **176**, 227 (2008).

[52] J. Jung and A. H. MacDonald, Accurate tight-binding models for the pi bands of bilayer graphene, Phys. Rev. B **89**, 035405 (2014).

[53] S. J. Yuan, R. Roldan, and M. I. Katsnelson, Landau level spectrum of ABA- and ABC-stacked trilayer graphene, Phys. Rev. B **84**, 125455 (2011).

[54] M. Koshino and T. Ando, Magneto-optical properties of multilayer graphene, Phys. Rev. B **77**, 115313 (2008).

**Figure captions**

**Figure 1.** (a) Optical image of trilayer graphene sheet. (b) Raman spectra of trilayer graphene taken at different domains with stacking order of ABA and ABC. (c) White light reflectance image of trilayer graphene sheet shown in panel (a). (d) and (e) are Raman images of G band intensity and G'(2D) band linewidth of trilayer graphene sheet, respectively.

**Figure 2.** False-colour map of the Raman intensity in the energy range of the $E_{2g}$ optical phonon as a function of the magnetic field for (a) ABA-stacked and (b) ABC-stacked trilayer graphene measured at $T$=4.2 K under λ=532 nm laser excitation. Magnetic field evolution of (c) G peak position, (d) G peak linewidth (FWHM), and (e) G peak height of ABA- and ABC-stacked trilayer graphene by fitting the Raman G band at different values of magnetic field as a single Lorentzian peak. (f), (g) and (h) are the amplitude of the Fourier transform of their corresponding insets, showing the G peak position, G peak linewidth, and G peak height of ABA-stacked trilayer graphene as a function of 1/B, respectively. We note that in sufficiently clean samples it is possible to observe clearly the anti-crossing branches that underlie the oscillatory features observed in our experiments.

**Figure 3.** *Left Panel:* Theoretical calculations of the magnetophonon oscillations for ABA- and ABC-stacked trilayer graphene. The oscillations for ABA-stacked TLG have been modeled using the dominant monolayer Landau levels in the system. In ABC-stacked TLG, we have used the low energy approximation for the Landau level spectrum. Because this approximation overestimates the Landau level spacing for moderately large magnetic fields we use a correction factor $\xi = 1/4$ which in turn leads to a suppression of the magnetophonon oscillations. We used the parameter values of $t = 0.4, 0.9$ eV for the ABC trilayer and $t = 1.2$ eV for ABA trilayer to control the vertical position of the magnetophonon oscillation curves. *Right Panel:* The first twenty Landau levels in monolayer graphene and ABC trilayer graphene. In the ABC trilayer case we plotted the full-bands Landau levels in blue and the corrected low energy form we used in our calculations. The comparison of the vertical axis energy scales between the monolayer and ABC trilayer graphene shows that the latter has more compactly bunched Landau levels due to its higher band chirality.

**Figure 4.** Raman spectra of (a) ABA-stacked and (b) ABC-stacked trilayer graphene at different values of magnetic field measured at $T$=4.2 K under $\lambda$=532 nm laser excitation.

**Figure 5.** (a) Optical image of suspended and supported trilayer graphene sheet. The Raman spectra and Raman images (not shown here) indicate that the whole piece is ABA-stacked trilayer graphene. (b) and (c) are magnetic field evolution of the G peak position and the G peak linewidth, respectively, of both suspended and supported ABA-stacked trilayer graphene.

**Figure 6.** (a) Optical and (b) white light reflectance images of trilayer graphene sheet. (c) and (d) are Raman images extracted from G band intensity and G'(2D) band linewidth, respectively. (e), (f) and (g) are magnetic field evolution of the G peak position, the G peak linewidth, and the G peak height, respectively, of ABC-stacked trilayer graphene.

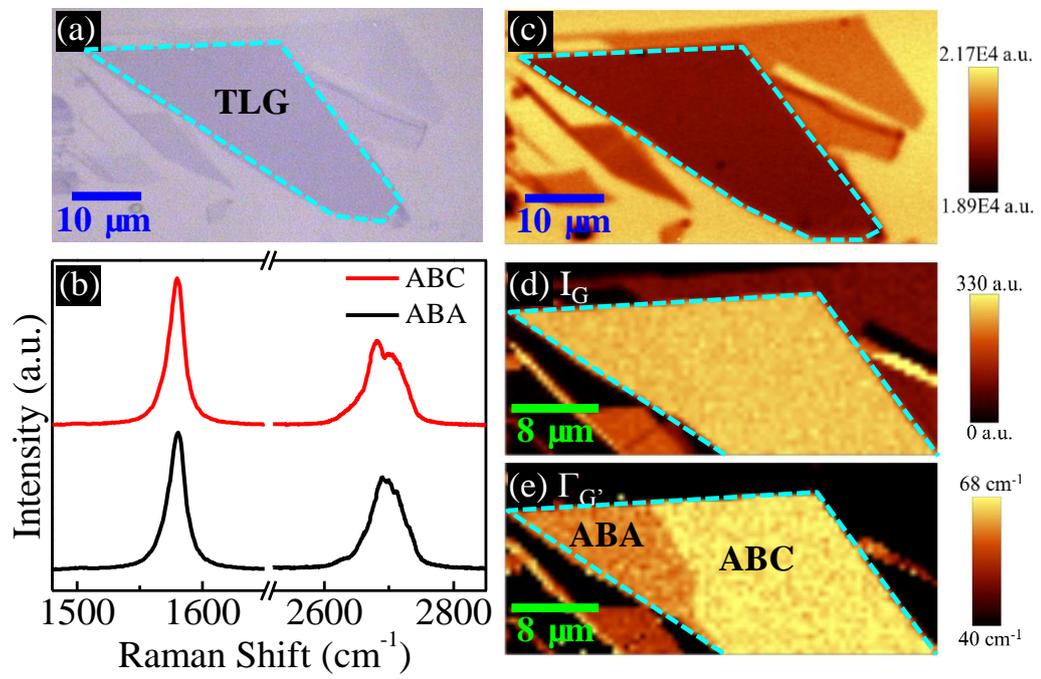

**Figure 1**

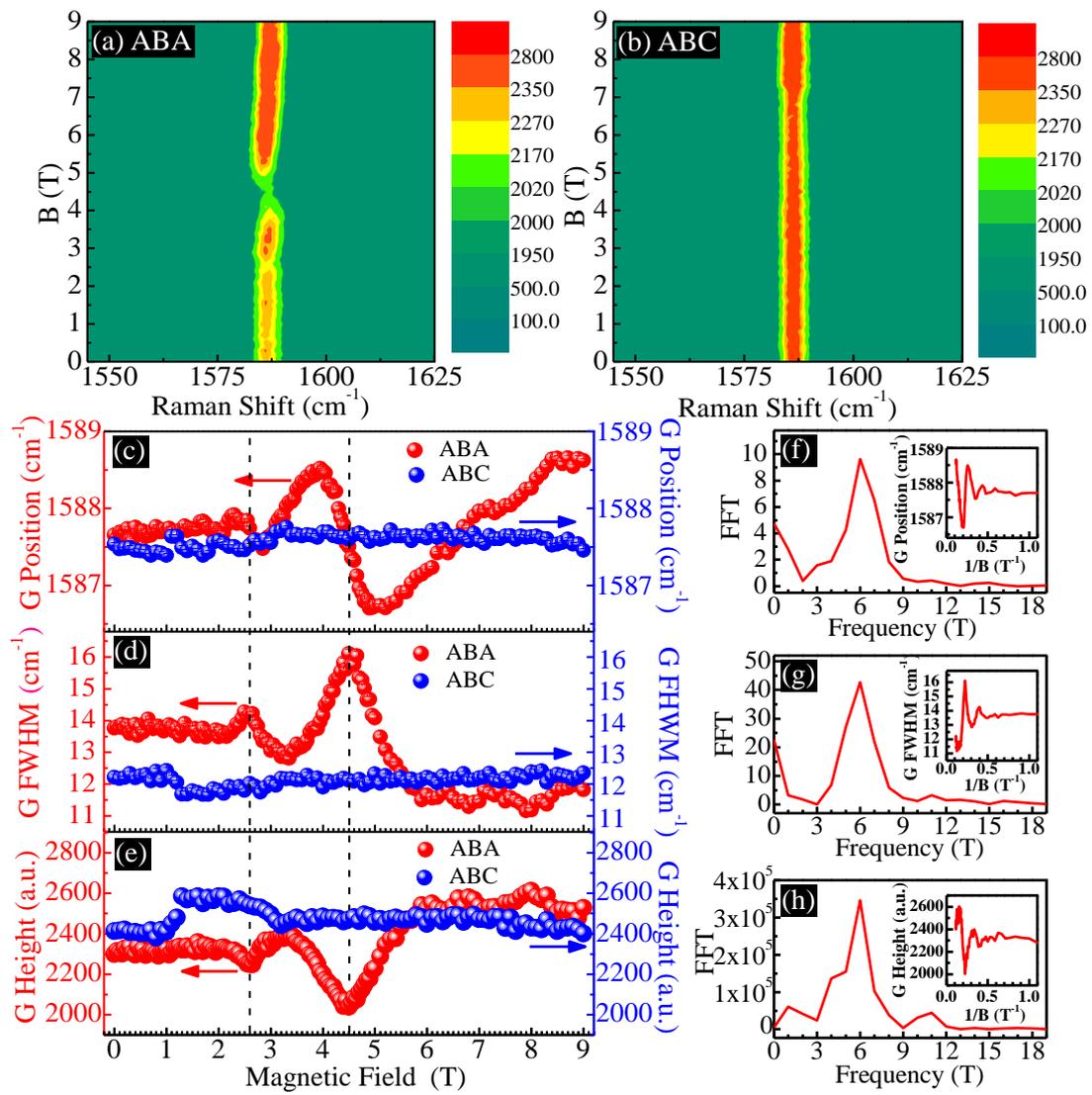

**Figure 2**

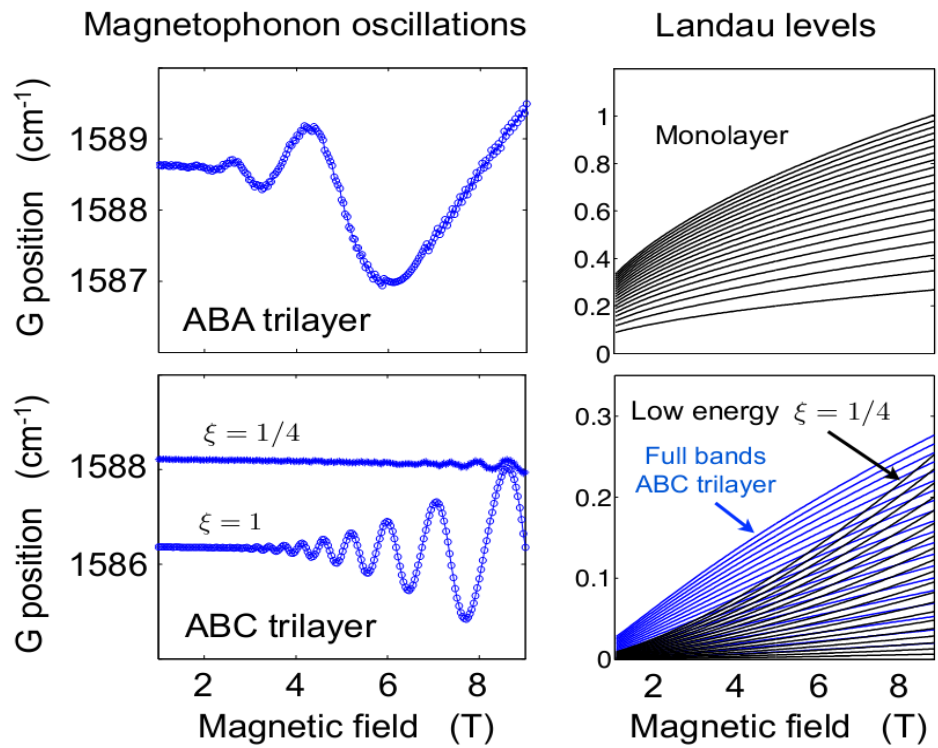

**Figure 3**

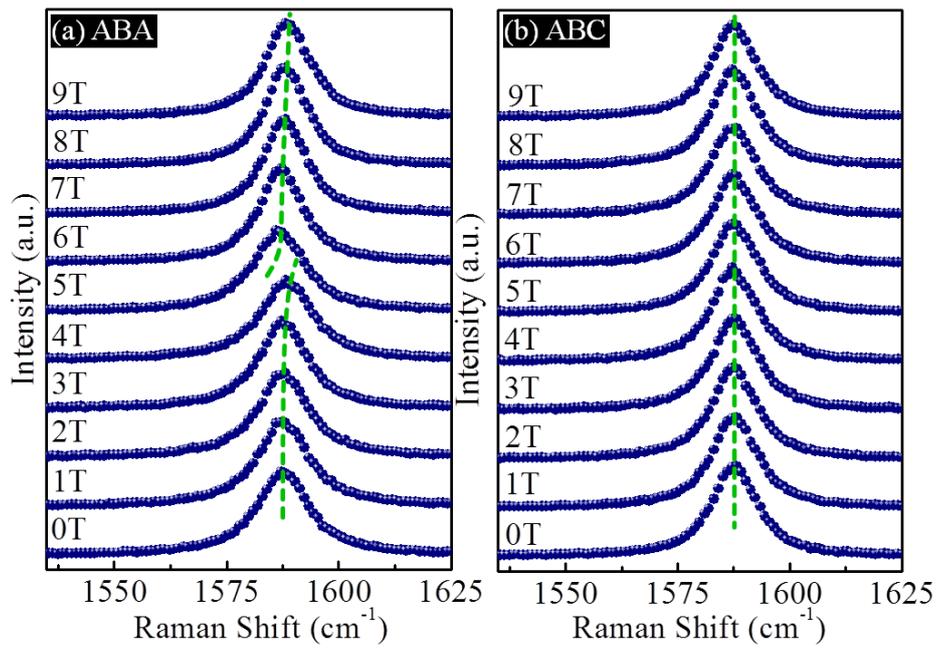

**Figure 4**

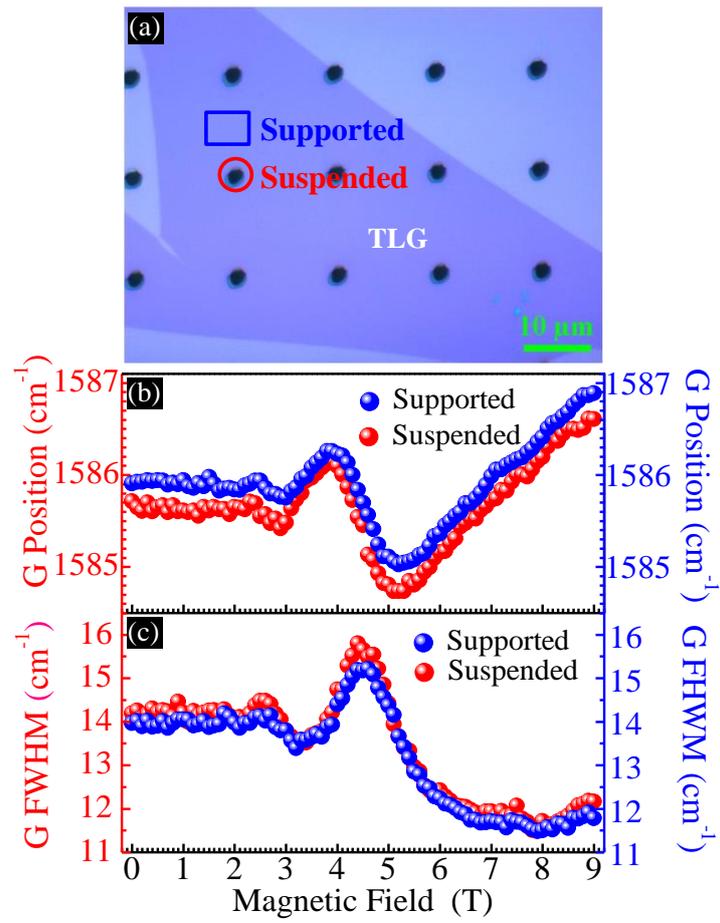

**Figure 5**

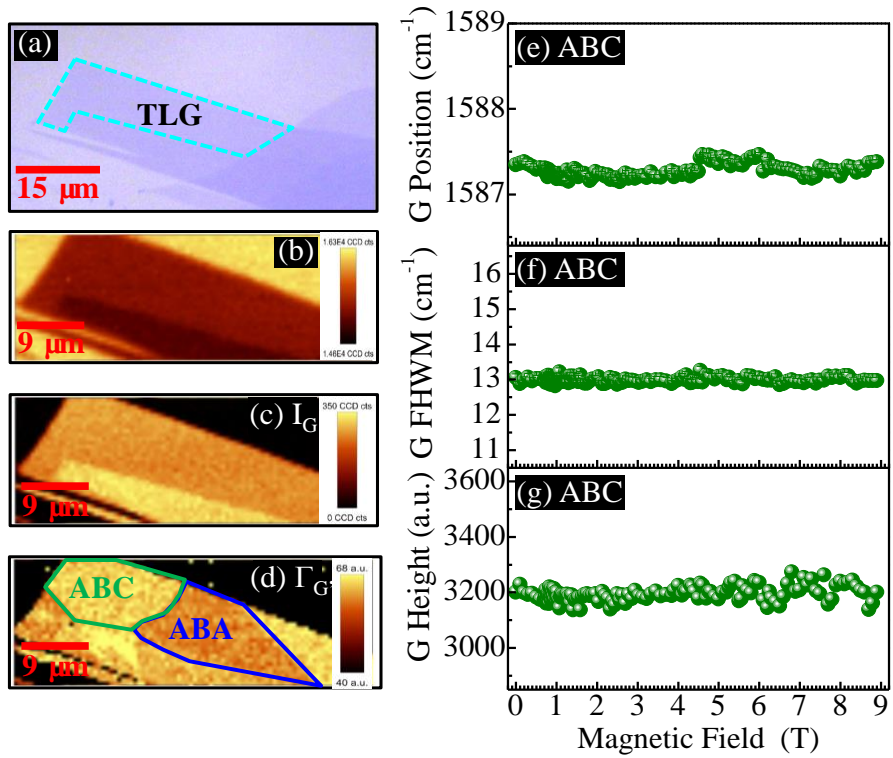

**Figure 6**